\documentclass{article}
\usepackage{latexsym,amssymb,amsmath}   
\usepackage[dvips]{graphicx}

\usepackage{latexsym,amsmath}
\usepackage{amssymb,amsthm}
\newcommand{\br}{\begin{eqnarray*}}
\newcommand{\er}{\end{eqnarray*}}

\newcommand{\ket}[1]{\arrowvert{#1}\rangle }

\newcommand{\bra}[1]{\langle{#1}\arrowvert}

\newtheorem{law}{Law}

\newtheorem{df}[law]{Definition}

\newtheorem{theorem}[law]{Theorem}

\newcommand{\eqdef}{\overset{\text{\begin{tiny}DEF\end{tiny}}}{=}}

\newcommand{\norm}[1]{\lvert{#1}\rvert}
\newcommand{\bignorm}[1]{\Bigl\lvert{#1}\Bigr\rvert}

\newcommand{\TM}{{Turing machine}}

\newcommand{\PTM}{{probabilistic Turing machine}}
\newcommand{\QTM}{{quantum Turing machine}}

\newcommand{\captionfonts}{\small\sl}

\makeatletter  % Allow the use of @ in command names
\long\def\@makecaption#1#2{%
  \vskip\abovecaptionskip
  \sbox\@tempboxa{{\captionfonts #1: #2}}%
  \ifdim \wd\@tempboxa >\hsize
    {\captionfonts #1: #2\par}
  \else
    \hbox to\hsize{\hfil\box\@tempboxa\hfil}%
  \fi
  \vskip\belowcaptionskip}
\makeatother   % Cancel the effect of \makeatletter

\begin{document}
\title{A new sibling of $BQP$}

\author{Tereza Tu\v sarov\' a\footnote{tusarova@qubit.cz, Charles University, Faculty  of  Mathematics  and  Physics, Prague, Czech Republic.}}
\date{10.4.2005}

\hyphenation
{ groun-ded de-tec-tion en-tan-gled ge-ne-rally phy-si-cally simi-larity defi-nition defi-nitions con-figurations exists}

\maketitle

\begin{abstract}
We present a new quantum complexity class, called MQ$^2$, which is contained in AWPP.
This class has a compact and simple mathematical definition, involving only polynomial-time computable functions and a unitarity condition. It contains both Deutsch-Jozsa's and Shor's algorithm, while
its relation to BQP is unknown. This shows that in the complexity class hierarchy, BQP is not an extraordinary isolated island, but has ''siblings'' which as well can solve prime-factorization.
\end{abstract}
%\keywords{quantum complexity, quantum algorithms}

\section{Introduction}
Quantum computing used to be a very popular discipline ranging from pure theoretical questions, concerning the complexity of the quantum polynomial-time class BQP, to practical concerns such as how to build a quantum computer. 
Nowadays, it seems a little bit that scientists are loosing their interest. Is it because all ''easy'' questions have been answered and what reamins is too hard or unintresting? We can find many unanswered questions considering complexity classes. 
For example, we know that BPP$\subseteq$BQP$\subseteq$AWPP. But questions whether is BQP equal to its ''father'' AWPP or
its ''son'' BPP have not been answered up to now. 
Here, we do not answer them either. Instead, we introduce a nontrivial ''brother'', which we call MQ$^2$. Surprisingly, this brother can also factorize long integers in polynomial time. Moreover, it has a very compact mathematical definition, which does not explicitly involve any physics.

The paper is organized as follows.
In section 2, we introduce the neccessary notation and definitions. In section 3, we briefly review classical polynomial-time classes definition. In the following section, we define the class MQ$^2$ itself. Fifth sections demontrates two quantum algorithms, Shor's and Deutsch-Jozsa's, in class MQ$^2$.

\section{Definitions and notation}
In this text, we frequently encounter matrices. All matrices here will have square shape. We will denote the element of a matrix $M$ in $i-$th row and $j-$th column as $\bra{i}M\ket{j}$, in accordance with the usual notation used in quantum computing,
because we think it makes the text more readable than writing $M_{i,j}$. 
The range of the indices will be $0..n-1$ where $n$ is the number of columns or rows.

We will define our new complexity class with use of matrices. To make the class uniform, we will want that all the matrices are constructed by the same algorithm. We formalize this in the following definition.

\begin{df}[Poly-computable matrix family]\label{family}
A sequence of matrices $T_1,T_2,\ldots$ is called a poly-computable matrix family if there exists a function 
$f:\mathbb{N}\times\mathbb{N}\times\mathbb{N}\to \mathbb{R}$ computable in time polynomial in
the length of all the arguments such that 
$$\forall i,j,n: f(i,j,n)=\bra{j}T_n\ket{i}.$$
\end{df}
Sometimes, we will drop the index $n$ when it will be clear from the context.

To show how the classical polynomial-time classes definitions correspond to our definition, we will start with a Turing machine and express its transition function as a transition matrix. A transition matrix is in fact a linear operator defined in a vector space spanned by configurations playing role of base vectors. 

\begin{df}[Configuration of a \TM]\index{configuration}
A configuration of a \TM\ is an ordered triple consisting of:
\begin{itemize}
\item[]
-the contents of the tape\\
-the current state\\
-the position of the head \footnote{We assume without loss of generality that the machine has only one tape.}
\end{itemize}
\end{df}
We emphasize here that the configuration as defined above contains also the content of the tape, which is not true for configurations as defined elsewhere. Without loss of generality, we will further assume that configurations are indexed and denoted by their indices.
A special position among the configurations has the family of initial configuration $I(x)$, which we allow to be dependent on 
$x$, but require to be computable in polynomial time. Again, we will drop the index $n$ when it will be clear which member of the family we mean. 
We will also need to be able to recognize accepting configurations. For this purpose, we will have a function $a(x,c)$, computable in polynomial time, where the first argument will be the input for the algorithm and $c$ is a configuration. The function $a(x,c)$ will return 1 iff the configuration $c$ is accepting (possibly depending on $x$) and 0 otherwise.

%*****************************************************************************************************% Transition matrix
%*****************************************************************************************************I
Now we are ready to jump to the notion of transition matrix. If a configuration $c_1$ leads to another configuration $c_2$ in the next step with probability $p$, there is $p$ on the position $\bra{c_2}T\ket{c_1}$, otherwise there is zero.
Because the tapes are of unbounded size, so is the matrix. However, if we
know that the time complexity of a \TM\ is  $T(n)$, we may for fixed $n$ have a finite matrix cutting the tapes at the distance $T(n)$ from the initial position on both sides. The size of the matrix for inputs of length $n$ is then 
$2^{O(T(n))}\times 2^{O(T(n))}$.
For a \PTM , the transition matrix is stochastic, e.g. every row sums up to 1.
Transition matrices naturally form a poly-computable matrix family, since for each pair $c_1$, $c_2$, the probability of going from one to another can be read from the description of the underlying \PTM , which is a finite object\footnote{It should be pointed out that for most poly-computable stochastic matrix families, no corresponding \PTM\ exists. The reason is that each \PTM\ has finite description of a bounded size, independent on the length of input, while 
in our definition \ref{family}, we allowed function $f$ to arbitrarily depend on $n$.}.

%*****************************************************************************************************
\section{Traditional complexity classes}
%*****************************************************************************************************
We will now briefly review classical complexity classes definitions. The common definitions of P, BPP, NP and PP involve  a \PTM\ and look at the accepting probability for each input\footnote{Obviously, the class P can be viewed as a special case of a probabilistic class, with probabilities either one or zero.}.

One step of a \PTM\ corresponds to multiplying the transition matrix with a vector representing the current configuration. 
Thus, instead of saying ''the probability of accepting on a configuration $I(x)$ after $S$ steps is $p$'', we may eqvivalently say 
''$\sum_{c:a(c,x)=1}\bra{c}T^{S}\ket{I(x)}=p$''. We will use this observation in the following definitions.

%*****************************************************************************************************% Klasicka definice
%*****************************************************************************************************I
\begin{df}[Polynomial time classes in matrix notation]\label{def2}
A language $L$ is in class C if there exists a polynomial $p(n)$ and a \PTM\ $M$ with transition matrix family $T_i$
and functions $I(x)$, $a(x,c)$ computable in polynomial time, such that for all $n$ and for all $x$ of length $n$:
\begin{center}
	\begin{tabular}{|l|r|r|r|r|}
	\hline
	complexity class C & P & NP & PP & BPP   \\
	\hline
	For $x\in L$,   $\sum_{c:a(c,x)=1} \bra{c}T_n^{p(n)}\ket{I(x)}$  
	& $=1$ 	& $>0$	 & $>\frac{1}{2}$ 		& $\geq \frac{2}{3}$ \\
	For $x\notin L$,   $\sum_{c:a(c,x)=1} \bra{c}T_n^{p(n)}\ket{I(x)}$ 
	& $=0$ 	& $=0$ 	& $\leq\frac{1}{2}$ 	& $ \leq\frac{1}{3}$ \\
	\hline
	\end{tabular}
\end{center}
where exactly one of the columns applies.\\
\end{df}

In the same manner, we may define the quantum class BQP:

\begin{df}[BQP]\label{bqpdef}
A language $L$ is in class BQP if there exists a polynomial $p(n)$ and a \QTM\ $M$ with transition matrix family $T_i$ of unitary matrices and functions $I(x)$, $a(x,c)$ computable in polynomial time, such that
\begin{itemize}
\item[]For $x\in L$:   $\bignorm{\sum_{c:a(c,x)=1} \bra{c}T^{p(n)}\ket{I(x)}}^2\geq \frac{2}{3}$ 
\item[]For $x\notin L$:   $\bignorm{\sum_{c:a(c,x)=1} \bra{c}T^{p(n)}\ket{I(x)}}^2\leq\frac{1}{3}$
\end{itemize}
\end{df}

We emphasize here that there are exactly two points in which Definition \ref{bqpdef} and the Definition of BPP in Definition \ref{def2} differ: First, in Definition \ref{def2} we have stochastic matrices while in Definition \ref{bqpdef} we have
unitary matrices.  Second, in Definition \ref{def2} we are looking at the value of $\sum_{c:a(c,x)=1}\bra{c}T^{p(n)}\ket{I(x)}$, while in Definition \ref{bqpdef} we look at the square norm of this value.  
However, the latter can be avoided, since we can eqvivalently use the square norm in the Definition \ref{def2} of BPP. 
Thus, the only remaining difference between the two classes is the type of matrices used.

Now we will define the class MQ$^2$ itself. 
\section{The definition}
We alter the Definition \ref{bqpdef} of class BQP to get a new class MQ$^2$.
At first, we will not use transition matrices, but instead a unitary, poly-computable matrix family. 
That is a weaker requirement. In turn, we will be more strict in the number of matrices allowed - we will only use two copies of the matrix.

%************************************************************************************************* 
% MQ^poly
%*************************************************************************************************
\newpage
\begin{df}[MQ$^2$]\label{mqdef}
A language $L$ is in class MQ$^2$ iff there exists a unitary, poly-computable matrix family $T$,
A poly-computable vector family $I(x)$ and function $a(x,c)$ computable in polynomial time such that
\footnote{
The numbers $\frac{2}{3}$ and $\frac{1}{3}$ in the definition can be amplified in the same way as in the case of BPP and BQP. Proven in \cite{vudipka}
}
\begin{itemize}
\item[]For $x\in L$,   $\sum_{c:a(c,x)=1}  \bignorm{\bra{c}T^2\ket{I(x)}}^2\geq \frac{2}{3}$ 
\item[]For $x\notin L$,   $\sum_{c:a(c,x)=1}\bignorm{\bra{c}T^2\ket{I(x)}}^2\leq\frac{1}{3}$ 
\end{itemize}
\end{df}

\begin{figure}[h]
\begin{center}
\includegraphics[width=220pt]{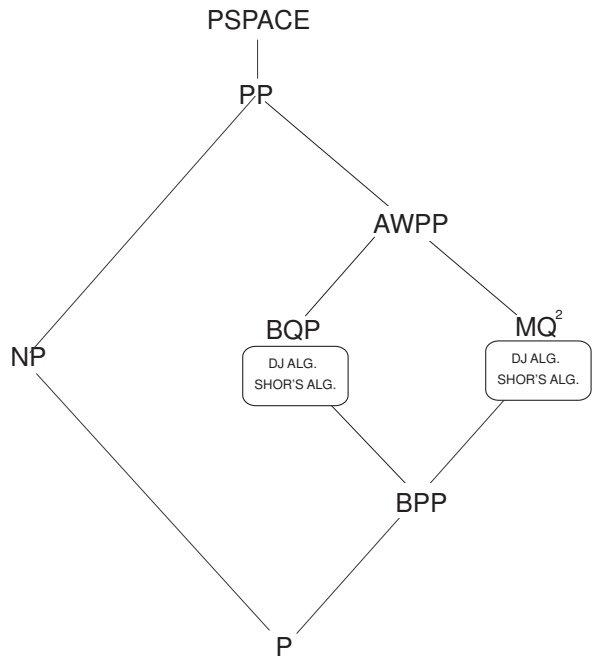}
\caption{Hierarchy of classes including MQ$^2$. For each pair connected by a line, the class that stays upper contains the lower one.}
\label{newclasses}
\end{center}
\end{figure}

The resulting hierarchy is visualised in Figure \ref{newclasses}. The inclusion MQ$^2\subseteq$AWPP was shown in \cite{vudipka}.

%*****************************************************************************************************
% DJ pro BQ 2 2
%*****************************************************************************************************
\section{Expressing quantum algorithms}
In this section, we will demonstrate how two famous quantum algorithms fit into the class MQ$^2$.  Here we will only present the main ideas of the proofs. Full proofs can be found in \cite{vudipka}.

At first, we show that class MQ$^2$ captures Deutsch-Jozsa's problem. For this problem, see \cite{dj}.
In the Deutsch-Jozsa problem, a~quantum oracle is used. That is a diagonal quantum gate, or in other words a diagonal unitary matrix, realizing the transformation $x\to (-1)^{f(x)}$. Here, our matrix will be a product of a poly-computable matrix and
this oracle. The result if then poly-computable too.

\begin{figure}
\begin{center}
\includegraphics[width=1.0\textwidth]{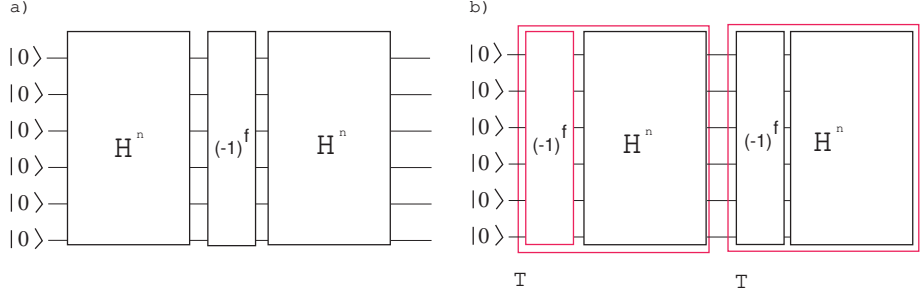}
\caption{The trick used to fit Deutsch-Jozsa's algorithm into the MQ$^2$ class. In a), there is the original setup,
in b) there is the one we use, which gives the same result up to a global phase. }
\label{dj}
\end{center}
\end{figure}

\begin{theorem}
The class MQ$^2$ solves the Deutsch-Jozsa problem. 
\end{theorem}
\noindent
\textit{Proof sketch. }We will mimic the circuit from Deutsch-Jozsa's algorithm (see Figure \ref{dj} a)) by two copies
of a poly-computable matrix family $T$ (see Figure \ref{dj} b)). For a fixed $n$, 
the matrix $T$ will be a product of the oracle and $H^n$.  We may add another matrix for the $f$ function to the front,
since it will only add the number $(-1)^{f(0)}$ to the global phase and thus will not change the result.
Formally, we define a matrix $T$ as

\begin{align*}
\bra{y}T\ket{x} \eqdef (-1)^{f(x)}\bra{y}H^n\ket{x}=\frac{1}{\sqrt{2^n}}(-1)^{f(x)}(-1)^{\sum_i x_i y_i \bmod 2}
\end{align*}
which is obviously computable in polytime and unitary.
We define $c_i(x)=0^n$ and $c_A(x)=0^n$ for $x$ of length $n$.
Then we have
\begin{multline*}
\bignorm{\bra{0^n}T^2\ket{0^n}}^2=\bignorm{\sum_k \bra{0^n}T\ket{k}\bra{k}T\ket{0^n}}^2=\\
\bignorm{\sum_k (-1)^{f(k)}\frac{1}{\sqrt{2^n}}(-1)^{\sum_i k_i {0^n}_i \bmod 2}\frac{1}{\sqrt{2^n}}(-1)^{f(0^n)}(-1)^{\sum_i {0^n}_i k_i \bmod 2}}^2=\\=
\bignorm{\sum_k \frac{1}{2^n}(-1)^{f(k)}(-1)^{2\sum_i k_i {0^n}_i \bmod 2}}^2=
\frac{1}{2^{2n}}\bignorm{\sum_k (-1)^{f(k)}}^2
\end{multline*}
If the function is constant, then the sum $\sum_k (-1)^{f(k)}$ equals $\pm 2^{n}$ and the
probability $\norm{\bra{0^n}T^2\ket{0^n}}^2$ equals $1$. If the function is balanced, both the sum and the probability is 0.
\begin{flushright}
$\Box$
\end{flushright}

%*****************************************************************************************************% Shor pro BQ2 2
%*****************************************************************************************************
Now we will show how to implement the famous Shor's algorithm \cite{shor} in class MQ$^2$:

\begin{theorem}
The class MQ$^2$ solves the factoring problem. More precisely,
there exists a constant k such that
given numbers x and N$\geq$k as in the  Shor's algorithm,
the language
\begin{equation*}
L=\{ \langle N,i\rangle | \text{$x^a \mod N$ has a period $r$ whose i-th bit is 1}\}
\end{equation*}
is in MQ$^2$.
\end{theorem}
\noindent
\textit{Proof sketch. }We will use the same notation as in the original paper by Shor
\cite{shor}. $N$ is the number to factorize and $N=p_1p_2$
where both $p_1$ $p_2$ are primes and are different from each
other. We then arbitrarily choose an $x$ coprime to $N$. The pair
$(x,N)$ is the input of the algorithm. The goal is to find the
smallest $r\neq 0$ such that $x^r\bmod N=1$. This $r$ is called a
\emph{period} of $x$. We choose a number $q$ such that $q$ is a
power of $2$ and $q\geq 2^{2\lceil log_2N \rceil}$. This number
will, together with the length of $x$ and $N$, determine the size of the
matrix.

\begin{figure}
\begin{center}
\includegraphics[width=1.0\textwidth]{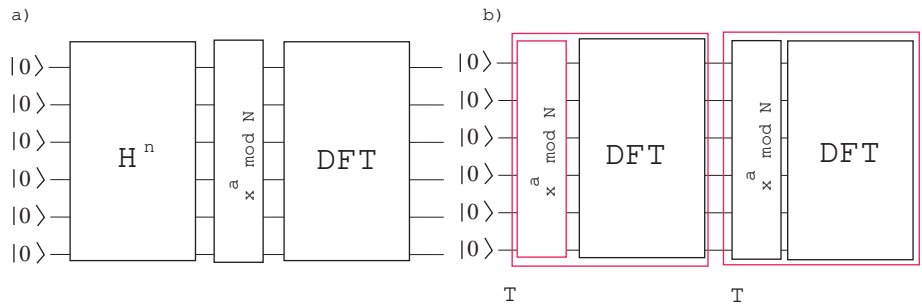}
\caption{The trick used to  fit the Shor's algorithm into MQ$^2$ class. In a), there is the original setup,
in b) there is the one we use, which gives the same result up to a global phase. }
\label{shor}
\end{center}
\end{figure}

In the original setup, see Figure \ref{shor} a), we have three different transformations, namely: Hadamard transformation, a transformation computing power modulo $N$, and the $DFT$. One can notice that the effect on $0^n$ of the $DFT$ and of the Hadamard transform is the same, so we suffice with only two different transformations. Furthermore, the matrices realizing 
them are poly-computable:
\begin{align*}
\bra{x',N',a',i'}DFT\ket{x,N,a,i}&\eqdef\frac{1}{\sqrt{q}}\delta_{x,x'}\delta_{N,N'}\delta_{i,i'}e^{\frac{2i\pi}{q}aa'}\\
\bra{x',N',a',i'}MOD\ket{x,N,a,i}&\eqdef\delta_{x,x'}\delta_{N,N'}\delta_{a,a'}\delta_{i'+i,x^a\bmod N}\\
\end{align*}
One may also simply check that both the matrices are unitary. Their product  $T\eqdef DFT\cdot MOD$ is thus unitary too. 
Its elements read
\begin{equation*}
\bra{x',N',a',i'}T\ket{x,N,a,i}=\frac{1}{\sqrt{q}}\delta_{x,x'}\delta_{N,N'}e^{\frac{2i\pi}{q}a'a}\delta_{i+i',x^a\bmod N}
\end{equation*}
and are again clearly functions computable in polytime.
Applying two times the matrix $T$, we apply an extra $MOD$ transformation comparing to the original setup. Nevertheless, on $0^n$, this transform acts as identity. Any classical post-processing, as in the original algorithm, can be incorporated into the function $a(x,c)$.
\begin{flushright}
$\Box$
\end{flushright}

\section{Conclusion}
We saw a complexity class which has a compact purely mathematical definition. In order to describe an algorithm, we suffice with 
three polynomial-time computable functions taking bit-strings as arguments: $f(i,j,n)$, $I(x)$, and $a(x,c)$. Even quite a complex algorithm, as the Shor's certainly is, can be  described on three lines. 
Further, the class MQ$^2$ shows that BQP is not the only possible class, lying in between BPP and AWPP, and not being trivially equal to either of them, which can do factorization and exponential speedup with oracles as in Deutsch-Jozsa's algorithm. 
\thispagestyle{plain}
\section*{Acknowledgment}
This material is partially based on the author's master thesis \cite{vudipka}, which was supervised by Harry Buhrman.
I would like to use this opportunity to thank him for motivating discussions.
\thispagestyle{plain}

\bibliographystyle{acm} 
\bibliography{../lite}

\end{document}